\documentclass[12pt]{article}

\usepackage{amsmath}
\usepackage{amssymb}
\usepackage{amsfonts}
\usepackage{latexsym}
\usepackage{color}
\usepackage[english]{babel}
\usepackage[utf8]{inputenc}
\usepackage{amssymb,latexsym,amsmath,amsfonts}
\usepackage{graphicx}
\usepackage{latexsym}

\catcode `\@=11 \@addtoreset{equation}{section}

\catcode `\@=12

\newcommand{\be}{\begin{equation}}
\newcommand{\en}{\end{equation}}
\newcommand{\bea}{\begin{eqnarray}}
\newcommand{\ena}{\end{eqnarray}}
\newcommand{\beano}{\begin{eqnarray*}}
\newcommand{\enano}{\end{eqnarray*}}
\newcommand{\bee}{\begin{enumerate}}
\newcommand{\ene}{\end{enumerate}}

\newcommand{\N}{\mathfrak N}

\newcommand{\mc}{\mathcal}

\newcommand{\Pc}{{\cal P}}

\newcommand{\F}{{\cal F}}

\newcommand{\T}{\cal{T}}
\newcommand{\1}{1 \!\! 1}

\newcommand{\Hil}{\mc H}
\newcommand{\kt}{\rangle}
\newcommand{\br}{\langle}

\catcode `\@=11 \@addtoreset{equation}{section}
\catcode `\@=12

\begin{document}

\thispagestyle{empty}

\vspace*{2cm}

\begin{center}
{\Large \bf Model pseudofermionic systems: connections with exceptional points}   \vspace{2cm}\\

{\large F. Bagarello}\\
  DEIM,
Facolt\`a di Ingegneria,\\ Universit\`a di Palermo, I-90128  Palermo, Italy and\\
\, INFN, Sezione di Torino, Italy\\
e-mail: fabio.bagarello@unipa.it\\
home page: www.unipa.it$\backslash$fabio.bagarello

\vspace{.8cm}

{\large F. Gargano}\\
 CNR,
IAMC,\\ Via Vaccara 61, I-91026  Mazzara del Vallo, Italy\\
e-mail: francesco.gargano@unipa.it\\

\end{center}

\vspace*{2cm}

\begin{abstract}
\noindent We discuss the role of pseudo-fermions in the analysis of some two-dimensional models, recently introduced in connection with non self-adjoint hamiltonians. Among other aspects, we discuss the appearance of exceptional points in connection with the validity of the extended anti-commutation rules which define the pseudo-fermionic structure.

\end{abstract}

\vspace{2cm}


\vfill


\newpage

\section{Introduction}

In recent years, extending what was previously done with canonical commutation relations, one of us (F.B.) considered a deformed version of the canonical anti-commutation relation (CAR), \cite{bagpf}, leading to an interesting functional structure: biorthogonal bases $\F_\varphi=\{\varphi_0,\varphi_1\}$ and $\F_\Psi=\{\Psi_0,\Psi_1\}$ appear, as well as lowering, raising and  not self-adjoint number operators $N$ and $N^\dagger$, whose eigenvectors are exactly the elements in $\F_\varphi$ and $\F_\Psi$. Also, we find intertwining operators connecting $N$ and $N^\dagger$ which are bounded, invertible, and self-adjoint. The same structure can be extended to more {\em pseudo-fermionic  modes}, and some applications to optical and electronic systems have also been proposed, \cite{bagpf2,bagpantano}.

Here we discuss systematically how the single-mode pseudo-fermions (PFs) can be naturally used, in the context of some models introduced in recent years in connection with pseudo-hermitian systems. Among other aspects, we consider exceptional points (EPs), trying to characterize them in terms of our modified CAR. Our main conclusion is that EPs are linked to the absence of PFs: in all the models considered here we will show that, in correspondence of their EPs, it becomes impossible to introduce operators satisfying the required anti-commutation rules, while, whenever these rules (see (\ref{220}) below) are satisfied, we are away from EPs.

The paper is organized as follows: in the next section we briefly discuss some basic facts on PFs. Section II.1 is devoted to a rather general construction, i.e. to the more general non self-adjoint hamiltonian which can be discussed in terms of pseudo-fermionic operators, whose symmetries are analyzed in Section II.2. In Section III we show how this general hamiltonian can be used in some recent examples of $2\times2$ non self-adjoint hamiltonians proposed by Bender, Jones, Mostafazadeh and others.  Our conclusions are given in
Section IV.

\section{Pseudo-fermions and hamiltonians}

We begin this section by briefly reviewing the main definitions and results concerning single-mode PFs. The extension to higher dimensions
will be discussed later on. The starting point is a modification of the CAR $\{c,c^\dagger\}=c\,c^\dagger+c^\dagger\,c=\1$,
$\{c,c\}=\{c^\dagger,c^\dagger\}=0$, between two operators, $c$ and $c^\dagger$, acting on a two-dimensional Hilbert space $\Hil$. The CAR are
replaced here by the following rules: \be \{a,b\}=\1, \quad \{a,a\}=0,\quad \{b,b\}=0, \label{220}\en where the interesting situation is when
$b\neq a^\dagger$. These rules automatically imply that a non zero vector, $\varphi_0$, exists in $\Hil$ such that $a\,\varphi_0=0$, and that a
second non zero vector, $\Psi_0$, also exists in $\Hil$ such that $b^\dagger\,\Psi_0=0$, \cite{bagpf}. In general $\varphi_0\neq\Psi_0$.

Let us now introduce the following non zero
vectors \be \varphi_1=b\varphi_0,\quad \Psi_1=a^\dagger \Psi_0, \label{221}\en as well as the non self-adjoint operators \be N=ba,\quad
\N=N^\dagger=a^\dagger b^\dagger. \label{222}\en We also introduce the self-adjoint operators $S_\varphi$ and $S_\Psi$ via their action on a
generic $f\in\Hil$: \be S_\varphi f=\sum_{n=0}^1\br\varphi_n,f\kt\,\varphi_n, \quad S_\Psi f=\sum_{n=0}^1\br\Psi_n,f\kt\,\Psi_n. \label{223}\en
Hence we get the following results,   whose proofs are straightforward and will not be given here:

\begin{enumerate}

\item \be a\varphi_1=\varphi_0,\quad b^\dagger\Psi_1=\Psi_0.\label{224}\en

\item \be N\varphi_n=n\varphi_n,\quad \N\Psi_n=n\Psi_n,\label{225}\en
for $n=0,1$.

\item If the normalizations of $\varphi_0$ and $\Psi_0$ are chosen in such a way that $\left<\varphi_0,\Psi_0\right>=1$,
then \be \left<\varphi_k,\Psi_n\right>=\delta_{k,n},\label{226}\en for $k,n=0,1$.

\item $S_\varphi$ and $S_\Psi$ are bounded, strictly positive, self-adjoint, and invertible. They satisfy
\be \|S_\varphi\|\leq\|\varphi_0\|^2+\|\varphi_1\|^2, \quad \|S_\Psi\|\leq\|\Psi_0\|^2+\|\Psi_1\|^2,\label{227}\en \be S_\varphi
\Psi_n=\varphi_n,\qquad S_\Psi \varphi_n=\Psi_n,\label{228}\en for $n=0,1$, as well as $S_\varphi=S_\Psi^{-1}$. Moreover, the following
intertwining relations \be S_\Psi N=\N S_\Psi,\qquad S_\varphi \N=N S_\varphi,\label{229}\en are satisfied.

\end{enumerate}

The above formulas show that (i) $N$ and $\N$ behave essentially as fermionic number operators, having eigenvalues 0 and 1 ; (ii) their related eigenvectors
are respectively the vectors of $\F_\varphi=\{\varphi_0,\varphi_1\}$ and $\F_\Psi=\{\Psi_0,\Psi_1\}$; (iii) $a$ and $b^\dagger$ are lowering
operators for $\F_\varphi$ and $\F_\Psi$ respectively; (iv) $b$ and $a^\dagger$ are rising operators for $\F_\varphi$ and $\F_\Psi$
respectively; (v) the two sets $\F_\varphi$ and $\F_\Psi$ are biorthonormal; (vi) the {\em well-behaved} (i.e. self-adjoint, bounded, invertible, with bounded inverse) operators $S_\varphi$ and
$S_\Psi$  maps $\F_\varphi$ in $\F_\Psi$ and viceversa; (vii) $S_\varphi$ and $S_\Psi$ intertwine between operators which are not self-adjoint.

We refer to \cite{bagpf} and \cite{bagpf2} for further remarks and consequences of these definitions. In particular, for instance, it is shown that $\F_\varphi$
and $\F_\Psi$ are automatically Riesz bases for $\Hil$, and the relations between fermions and PFs are also discussed.

Going back to (\ref{220}), as we have discussed in \cite{bagpf}, the only non-trivial possible choices of $a$ and $b$ satisfying these rules
are the following:
$$
a(1)=\left(
       \begin{array}{cc}
         0 & 1 \\
         0 & 0 \\
       \end{array}
     \right), \quad b(1)=\left(
                           \begin{array}{cc}
                             \beta & -\beta^2 \\
                             1 & -\beta \\
                           \end{array}
                         \right), $$ $$a(2)=\left(
       \begin{array}{cc}
        \alpha & 1 \\
         -\alpha^2 & -\alpha \\
       \end{array}
     \right), \quad b(2)=\left(
                           \begin{array}{cc}
                             0 & 0 \\
                             1 & 0 \\
                           \end{array}
                         \right),
$$
with non zero $\alpha$ and $\beta$, or, maybe more interestingly,
$$
a(3)=\left(
       \begin{array}{cc}
        \alpha_{11} & \alpha_{12} \\
         -\alpha_{11}^2/\alpha_{12} & -\alpha_{11} \\
       \end{array}
     \right), \quad b(3)=\left(
       \begin{array}{cc}
        \beta_{11} & \beta_{12} \\
         -\beta_{11}^2/\beta_{12} & -\beta_{11} \\
       \end{array}
     \right),
$$
with \be 2\alpha_{11}\beta_{11}-\frac{\alpha_{11}^2\beta_{12}}{\alpha_{12}}-\frac{\beta_{11}^2\alpha_{12}}{\beta_{12}}=1.\label{230}\en Other possibilities
also exist, but are those in which $a$ and $b$ exchange their roles or those in which $a$ and $b$ are standard fermion operators. Also, these matrices are not
really all independent, since $a(1)$ and $b(1)$ can be recovered from  $a(3)$ and $b(3)$ taking $\alpha_{11}=0$,
$\alpha_{12}=1$, $\beta_{11}=\beta$, $\beta_{12}=-\beta^2$. Notice that this choice satisfies (\ref{230}).
Less trivially, we can also recover  $a(2)$ and $b(2)$ from  $a(3)$ and $b(3)$. In this case we need to take $\alpha_{11}=\alpha$, $\alpha_{12}=1$, $\beta_{11}=x$, $\beta_{12}=-x^2$, and then to send $x$ to zero. This means that, in order to consider the more general situation, it is enough to use the operators $a(3)$ and $b(3)$, endowed with condition (\ref{230}). From now on, this will be our choice, and we will simply write them $a$ and $b$.

\vspace{2mm}

{\bf Remark:--} For completeness we have to mention the paper by Bender and Klevansky, \cite{ben3}, where similar generalized anti-commutation rules were introduced, but with a different perspective.

\subsection{The hamiltonian}

In view of what we have just seen, the most general diagonalizable hamiltonian which can be written in terms of $a$ and $b$ is obviously the operator
\be
H=\omega N+\rho\1=\left(
                    \begin{array}{cc}
                      \omega\gamma\alpha+\rho & \omega\gamma \\
                      -\omega\gamma\alpha\beta & -\omega\gamma\beta+\rho \\
                    \end{array}
                  \right),
\label{231}\en
where $\omega$ and $\rho$, in principle, could be complex numbers,
$\alpha=\frac{\alpha_{11}}{\alpha_{12}}$, $\beta=\frac{\beta_{11}}{\beta_{12}}$,
and $\gamma=\alpha_{12}\beta_{11}-\alpha_{11}\beta_{12}=\alpha_{12}\beta_{12}(\beta-\alpha)$. Then we can write
$$
a=\alpha_{12}\left(
               \begin{array}{cc}
                 \alpha & 1 \\
                 -\alpha^2 & -\alpha \\
               \end{array}
             \right), \qquad b=\beta_{12}\left(
               \begin{array}{cc}
                 \beta & 1 \\
                 -\beta^2 & -\beta \\
               \end{array}
             \right),
$$
while condition (\ref{230}) can be written as $-\gamma^2=\alpha_{12}\beta_{12}$. This also implies that $(\alpha-\beta)\gamma=1$.

The eigensystem of $H$ is trivially deduced:
the eigenvalues are $\epsilon_0=\rho$ and $\epsilon_1=\omega+\rho$, which are real if and only if $\rho$ and $\omega$ are both real.
In this case, $\epsilon_0$ and $\epsilon_1$ are also the eigenvalues of $H^\dagger=\omega N^\dagger+\rho\1$. From now on, except when explicitly stated, we will assume that $\epsilon_j\in\Bbb R$, for $j=0,1$. It might be interesting to notice that, adopting the same limiting procedure described above ($\alpha_{11}=\alpha$, $\alpha_{12}=1$, $\beta_{11}=x$, $\beta_{12}=-x^2$, and $x\rightarrow0$), we simply recover $H=\rho\1$.

 The  eigenvectors of $N$ and $N^\dagger$, and of $H$ and $H^\dagger$ as a consequence, are the following:
\bea
\varphi_0=N_\varphi\left(
                     \begin{array}{c}
                       1 \\
                       -\alpha \\
                     \end{array}
                   \right),\qquad \varphi_1=b\varphi_0=\frac{\gamma N_\varphi}{\alpha_{12}}\left(
                     \begin{array}{c}
                       1 \\
                       -\beta \\
                     \end{array}
                   \right),\label{eigen1}\ena
 and                  \bea\Psi_0=N_\Psi\left(
                     \begin{array}{c}
                       1 \\
                       \overline{\beta}^{-1} \\
                     \end{array}
                   \right),\quad \Psi_1=a^\dagger\Psi_0=\frac{\overline{\gamma}\, N_\Psi}{\overline{\beta_{11}}}\left(
                     \begin{array}{c}
                       \overline{\alpha} \\
                       1 \\
                     \end{array}
                   \right),
\label{eigen2}\ena
where $N_\varphi\overline{N_\Psi}=\frac{\alpha_{12}\beta_{11}}{\gamma}$. This choice is dictated by the fact that $\left<\Psi_0,\varphi_0\right>=1$. Let us remind that $\varphi_0$ and $\Psi_0$ are (almost) fixed by requiring that they are annihilated by $a$ and $b^\dagger$, respectively: $
a\varphi_0=0$ and  $b^\dagger\Psi_0=0$.
Moreover we have $N\varphi_j=j\varphi_j$ and $N^\dagger\Psi_j=j\Psi_j$, $j=0,1$, so that
\be
H\varphi_j=\epsilon_j\varphi_j,\qquad H^\dagger\Psi_j=\epsilon_j\Psi_j,
\label{232}\en
$j=0,1$. Sometimes it can be useful to write $H$ and $H^\dagger$ in terms of the projectors $P_j$ defined as $P_jf=\left<\Psi_j,f\right>\varphi_j$, $j=0,1$, whose adjoint is $P_j^\dagger f=\left<\varphi_j,f\right>\Psi_j$ clearly\footnote{Of course they are not orthogonal projectors, since they are not self-adjoint, in general, and not even idempotent.}. Here $f$ is a generic vector in $\Hil$. Then $H=\epsilon_0P_0+\epsilon_1P_1$ and $H^\dagger=\epsilon_0P_0^\dagger+\epsilon_1P_1^\dagger$.

It is a straightforward computation to check that $\F_\varphi$ and $\F_\Psi$ produce, together, a resolution of the identity. Indeed we have $P_0+P_1=P_0^\dagger+P_1^\dagger=\1$. Hence, as expected, $\F_\varphi$ and $\F_\Psi$ are biorthogonal bases for $\Hil$.

The next step consists in finding the explicit expressions for $S_\varphi$ and $S_\Psi$ in (\ref{223}): we find
\be
S_\varphi=|N_\varphi|^2\left(
            \begin{array}{cc}
              1+\left|\frac{\gamma}{\alpha_{12}}\right|^2 & -\overline{\alpha}-\overline{\beta}\,\left|\frac{\gamma}{\alpha_{12}}\right|^2 \\
              -{\alpha}-{\beta}\,\left|\frac{\gamma}{\alpha_{12}}\right|^2  & |\alpha|^2+\left|\frac{\gamma\beta}{\alpha_{12}}\right|^2 \\
            \end{array}
          \right)
\label{add1}\en
and
\be
S_\Psi=|N_\Psi|^2\left(
            \begin{array}{cc}
              1+\left|\frac{\alpha\gamma}{\beta_{11}}\right|^2 & \frac{1}{\beta}+\overline{\alpha}\,\left|\frac{\gamma}{\beta_{11}}\right|^2 \\
              \frac{1}{\overline{\beta}}+{\alpha}\,\left|\frac{\gamma}{\beta_{11}}\right|^2  & \left|\frac{1}{\beta}\right|^2+\left|\frac{\gamma}{\beta_{11}}\right|^2 \\
            \end{array}
          \right),
\label{add2}\en
 which are both clearly self-adjoint\footnote{Notice that, since $\beta_{11}=\beta_{12}\beta$, we could rewrite $S_\Psi$ using $\beta_{12}$ rather than $\beta_{11}$. This could be useful in the following.}. Using, for instance, the Sylvester's criterion, it is possible to check explicitly that, if $\alpha\neq\beta$, both $S_\varphi$ and $S_\Psi$ are positive definite. This can also be deduced looking at the eigenvalues of the two matrices, or just using the definition: $\left<f, S_\varphi f\right>$ and  $\left<f, S_\Psi f\right>$
are both strictly positive for any non zero $f\in\Hil$, if $\alpha\neq\beta$. Interestingly enough,  $\alpha=\beta$ implies that condition (\ref{230}) cannot be satisfied, and this means, in turn, that we are loosing the pseudo-fermionic structure described before. In fact, $a$ and $b$ cannot satisfy any longer the anti-commutation rules in (\ref{220}). Therefore, it is not surprising that $S_\varphi$ and $S_\Psi$ do not admit inverse, contrarily to what happens whenever (\ref{220}) are satisfied. We will get a similar conclusion in explicit models: whenever $\alpha$ and $\beta$ coincide, our operators cannot satisfy (\ref{230}), or its equivalent expressions,  and PFs do not appear.

Because of their positivity, there exist unique square root matrices $S_\varphi^{1/2}$ and $S_\Psi^{1/2}$, which are also positive and self-adjoint. They have a rather involved expression, which we give here for just for completeness, but which is rather hard to manage:
\bea
S_\varphi^{1/2}=\frac{|N_{\phi}|}{\sqrt{2p_1}}\left(
\begin{array}{cc}
 \frac{\sqrt{p_3} \,p_5-\sqrt{p_2}\, p_4}{2 } & \overline{p}\\
p & \frac{\sqrt{p_2}\, p_5-\sqrt{p_3}\, p_4}{2 }
\end{array}
\right),\label{add12}\ena
and
\bea
S_\Psi^{1/2}=\frac{1}{| N_{\phi}|\sqrt{2q_1}}\left(
\begin{array}{cc}
 \frac{\sqrt{q_2}\, q_5-\sqrt{q_3}\, q_4}{2} & \overline{q}\\
 q & \frac{ \sqrt{q_3}\,q_5- \sqrt{q_2}\,q_4}{2}
\end{array}
\right),\label{add22}
\ena
where we have defined the following quantities:
\beano\left\{
\begin{array}{ll}
p_1=(1+t-|\alpha |^2-t|\beta|^2)^2+
4 |\alpha +t\beta|^2, \\
p_2=1-\sqrt{p_1} +t +|\alpha |^2+t |\beta |^2,\\
p_3=1+\sqrt{p_1} +t +|\alpha |^2+t |\beta |^2,\\
p_4=1-\sqrt{p_1}+t-|\alpha |^2-t |\beta |^2,\\
p_5=1+\sqrt{p_1}+t-|\alpha |^2-t |\beta |^2,\\
p=(\sqrt{p_2}-\sqrt{p_3})(\alpha+t\beta),\\
q_1=(|\beta|^2+|\alpha_{11}|^2-1-|\alpha_{12}|^2)^2+4|\beta+\alpha|\alpha_{12}||^2,\\
q_2=1-\sqrt{q_1}+|\alpha_{11}|^2+|\alpha_{12}|^2+|\beta |^2,\\
q_3=1+\sqrt{q_1}+|\alpha_{11}|^2+|\alpha_{12}|^2+|\beta |^2,\\
q_4=1-\sqrt{q_1}+|\alpha_{11}|^2-|\alpha_{12}|^2-|\beta |^2,\\
q_5=1+\sqrt{q_1}+|\alpha_{11}|^2-|\alpha_{12}|^2-|\beta |^2,\\
q=(\sqrt{q_3}-\sqrt{q_2})(\beta+\alpha|\alpha_{12}|^2),
\end{array}
\right.
\enano
and where $t=\left|\frac{\gamma}{\alpha_{12}}\right|^2$. Other results which can be explicitly derived are the following:
\begin{enumerate}

\item $S_\varphi\Psi_n=\varphi_n$ and $S_\Psi\varphi_n=\Psi_n$, $n=0,1$;

\item  $S_\Psi N=N^\dagger  S_\Psi$ and  $S_\varphi N^\dagger = N S_\varphi$;

\item calling $c=S_\Psi^{1/2}a\,S_\Psi^{-1/2}$ we find that $c^\dagger=S_\Psi^{1/2}b\,S_\Psi^{-1/2}$, and that $\{c,c^\dagger\}=\1$, $c^2=0$;

\item calling $N_0=c^\dagger c$ we have $N_0=S_\Psi^{1/2}N\,S_\Psi^{-1/2}=S_\Psi^{-1/2}N^\dagger\,S_\Psi^{1/2}$;

\item $e_0=S_\Psi^{1/2}\varphi_0$ and $e_1=S_\Psi^{1/2}\varphi_1$ are eigenstates of $N_0$, with eigenvalues 0 and 1. Therefore, they are also eigenstates of the self-adjoint hamiltonian $h=S_\Psi^{1/2} H\,S_\varphi^{1/2}=\omega N_0+\rho\1$, with eigenvalues $\epsilon_0$ and $\epsilon_1$. The set $\{e_0,e_1\}$ is an orthonormal basis for $\Hil$.

\end{enumerate}

All these results are consequences of the pseudo-fermionic anticommutation rules in (\ref{220}), and have been deduced and analyzed in \cite{bagpf}-\cite{bagpantano}.

\subsection{Symmetry of the hamiltonian}

We continue our analysis of $H$ looking for some non-trivial two-by-two matrix $X$ which commutes with $H$. Of course,
not to make the situation trivial, we assume here that $\omega\neq0$. Otherwise $H=\rho\1$ and  $[H,X]=0$ for
each matrix $X$. This also happens when $\gamma=0$, i.e. when $\alpha=\beta$ (not necessarily zero). We recall
that, in this last case, we lose the rules in (\ref{220}), so that we are no longer dealing with PFs.
This is not a big surprise, since also in this case $H$ turns out to be just a multiple of the identity operator, so that each non zero vector of $\Hil$
is an eigenstate of $H$ with eigenvalue $\rho$.

In case $\omega$ and $\gamma$ are both non zero, $X=\left(
                    \begin{array}{cc}
                      x_{11} & x_{12} \\
                      x_{21} & x_{22} \\
                    \end{array}
                  \right)$
commutes with $H$ only if the following is true:

 $$ x_{21}=-x_{12}\alpha\beta, \quad x_{22}=x_{11}-x_{12}( \alpha +\beta),$$
where $x_{11}$ and $x_{12}$ are free parameters.

Moreover, if we also ask that $X^2=\1$, we should further require that
$$
x_{11}=-x_{22}=\frac{\alpha+\beta}{\alpha-\beta},\quad x_{12}=\frac{2}{\alpha-\beta},\quad x_{21}=-\frac{2\alpha\beta}{\alpha-\beta}.
$$
Of course, with these choices, also $Y=-X$ commutes with $H$ and satisfies $Y^2=\1$.

The matrix $X$ can be seen essentially as a generalized version of the $ \Pc \mathcal{T}$-symmetry, where
\bea
\Pc=\left(
  \begin{array}{cc}
    0 & 1 \\
    1 & 0 \\
  \end{array}
\right) \quad \mathcal{T} := \text{complex conjugate}.\label{PT}
\ena

The Hamiltonian $H$ in \eqref{231} is not generally $\Pc \mathcal{T}$-symmetric,
since the condition $[\Pc \mathcal{T},H]=0$ is not guaranteed in general. However, $H$ is $\Pc \mathcal{T}$-symmetric under the following conditions:
 \bea
\rho + \alpha\gamma\omega=\overline{\rho -\beta\gamma\omega}, \quad
\alpha\beta\gamma\omega=-\overline{\gamma\omega},
\label{PTcond}
\ena
and, in this case, the hamiltonian $H$ becomes
\be
H=\left(
                    \begin{array}{cc}
                      \omega\gamma\alpha+\rho & \omega\gamma \\
                      \overline{\omega\gamma} & \overline{\omega\gamma\alpha+\rho} \\
                    \end{array}
                  \right).
\label{HPT}\en

Here it is more convenient to rewrite its eigenvalues $\epsilon_0$ and $\epsilon_1$  as
 $\epsilon_{\pm}=\Re(\rho+\alpha\gamma\omega)\pm\sqrt{Q}$, $Q=|\gamma\omega|^2-(\Im(\rho+\alpha\gamma\omega))^2$,  and the relative  eigenvectors $\varphi_0, \varphi_1$ in \eqref{eigen1} as
\beano|\epsilon_{+}\rangle=\left( \begin{array}{c}\frac{i\Im(\rho+\alpha\gamma\omega)+\sqrt{Q}}{\overline{\gamma\omega}} \\ 1 \end{array}\right)=
\left( \begin{array}{c}-\beta^{-1} \\ 1 \end{array} \right), \quad
 |\epsilon_{-}\rangle=\left(\begin{array}{c}\frac{i\Im(\rho+\alpha\gamma\omega)-\sqrt{Q}}{\overline{\gamma\omega}} \\ 1 \end{array}\right)=
 \left( \begin{array}{c}-\alpha^{-1} \\ 1 \end{array}\right),
 \enano
  with an obvious notation and
 with an appropriate choice of normalization.
The analytic expression for $\epsilon_{\pm}$ shows that the eigenvalues of $H$ can either
be real or form complex conjugate pair according to the sign of $Q$.


The $\Pc \T$-symmetry is $unbroken$ for $|\gamma\omega|>|\Im(\rho+\alpha\gamma\omega)|$ and in this case
$$
\Pc \mathcal{T} |\epsilon_+ \rangle=\lambda_+|\epsilon _+ \rangle, \quad \Pc\mathcal{T} | \epsilon_{-}\rangle=\lambda_-|\epsilon_-\rangle,$$
with $\lambda_{\pm}=\frac{\overline{\gamma\omega}}{i\Im(\rho+\alpha\gamma\omega)\pm\sqrt{Q}}$. Notice that  $|\lambda_{\pm}|=1$, and therefore all the components
of the eigenvectors $|\epsilon_{\pm}\rangle$ have unitary modulus. This implies that $|\alpha|=|\beta|=1$. We recall the the eigenvalues of $H$ are actually $\rho$ and $\rho+\omega$, and
therefore the $unbroken$ $\Pc\mathcal{T}$-symmetry is only compatible with the condition that $\rho$ and $\omega$ are both reals.

For $|\gamma\omega|<|\Im(\rho+\alpha\gamma\omega)|$
the eigenvalues of $H$ become complex conjugates and the symmetry is $broken$ because
$$\Pc \mathcal{T} |\epsilon_+\rangle=\tilde\lambda_+|\epsilon_-\rangle,\quad \Pc\mathcal{T} | \epsilon_{-} \rangle=\tilde\lambda_-|\epsilon_+\rangle$$
with $\tilde\lambda_{\pm}=-i\frac{\overline{\gamma\omega}}{\Im(\rho+\alpha\gamma\omega)\mp\sqrt{\tilde Q}}$ and $\tilde Q=-Q$.
In this case   $|\tilde\lambda_{\pm}|=1$ and moreover $\alpha\overline{\beta}=1$. The presence of a pair of complex conjugate eigenvalues of
$H$ implies necessarily that
$\rho$ is imaginary and $\omega=-2i\Im(\rho)$.
For $|\gamma\omega|=|\Im(\rho+\alpha\gamma\omega)|$ an EP occurs. The eigenvalues coalesce to the real value
$\epsilon=\Re(\rho+\alpha\gamma\omega)=\Re(\overline{\rho-\beta\gamma\omega})=\rho$ and $|\epsilon_{+}\rangle=|\epsilon_{-}\rangle$ which, in turn,
implies that $\alpha=\beta$ so that $\gamma=0$
(we do not consider here the trivial case $\omega=0$):
in this case the conditions \eqref{230} is not satisfied, and no PFs exist. The formation of an EP is therefore related not only to
the absence of the imaginary part of $\rho$ but also to the non-existence of PFs.\\

Going back to our matrix $X$ above, it has not, as stated, the structure of a $\Pc \mathcal{T}$ operator, meaning with this that, even if $[X,H]=[\Pc \mathcal{T},H]=0$, $X$ cannot be identified with $\Pc \mathcal{T}$.
This is not a major problem since in the literature, see for instance
\cite{das,deng},
 extended versions  of $\Pc \mathcal{T}$-symmetry exist, where it is not required that
 $[\Pc,\mathcal{T}]=0$ or that $\Pc=\Pc^{\dagger}$. One of such an extension has the form
\bea
\tilde\Pc=\left(
  \begin{array}{cc}
    0 & x \\
    1/x & 0 \\
  \end{array}
\right), \label{PTs}
\ena
with $x\neq0$. If we take $x$ real, the $\tilde \Pc \mathcal{T}$-symmetry condition $[\tilde \Pc \mathcal{T},H]=0$ is satisfied for the following conditions:
 \bea
\rho + \alpha\gamma\omega=\overline{\rho -\beta\gamma\omega}, \quad
x^2\alpha\beta\gamma\omega=-\overline{\gamma\omega},
\label{PT2cond}
\ena
which extend those in \eqref{PTcond}. It is possible to generalize our previous results to this situation: in fact taking
into account \eqref{PT2cond} the eigenvalues of $H$ are
 $\epsilon_{x_{\pm}}=\Re(\rho+\alpha\gamma\omega)\pm x^{-2}\sqrt{Q_x}$, and the relative  eigenvectors are
\beano
|\epsilon_{x_{+}}\rangle=\left( \begin{array}{c}\frac{ix^2\Im(\rho+\alpha\gamma\omega)+\sqrt{Q_x}}{\overline{\gamma\omega}} \\ 1 \end{array}\right), \quad
|\epsilon_{x_{-}}\rangle=\left(\begin{array}{c}\frac{ix^2\Im(\rho+\alpha\gamma\omega)-\sqrt{Q_x}}{\overline{\gamma\omega}} \\ 1 \end{array}\right),
 \enano
 where $Q_x=x^2|\gamma\omega|^2-x^4(\Im(\rho+\alpha\gamma\omega))^2$. For $Q_x>0$  we are in the domain of the $unbroken$ $\tilde \Pc \mathcal{T}$-symmetry, and the condition $|\alpha|=|\beta|=x^{-2}$ holds.
 The $broken$ $\Pc \mathcal{T}$-symmetry occur for $Q_x<0$, and in this case $\overline{\alpha}\beta=x^{-2}$ holds.
 An EP occur for  $Q_x=0$, i.e when $|\gamma\omega|=x^2|\Im(\rho+\alpha\gamma\omega)|$, and as in the specifc case of the $\Pc \mathcal{T}$-symmetry,
 the eigenvalues coalesce to $\epsilon_{x}=\rho$ and $|\epsilon_{x_{+}}\rangle=|\epsilon_{x_{-}}\rangle$, which implies that $\alpha=\beta$ with $\gamma=0$.
 This condition is again incompatible with the existence of pseudo fermions because \eqref{230} is no more verified .

\section{Examples from literature}

In this section we show how the above general framework can be used in the analysis of several concrete models introduced along the years by several authors. In other words, we will see that many simple systems considered by many authors fit very well into our framework.

\subsection{An example by Das and Greenwood}

The first example we want to consider was originally discussed, in our knowledge, in \cite{das}, and, in a slightly different version, by others. The hamiltonian is
\be
H_{DG}=\left(
  \begin{array}{cc}
    r\, e^{i\theta} & s\,e^{i\phi} \\
    t\,e^{-i\phi} & r\, e^{-i\theta} \\
  \end{array}
\right),
\label{m1}\en
where $r, s, t, \theta$ and $\phi$ are all real quantities. In particular, to make the situation more interesting, we will assume that $r, s$ and $t$ are non zero. We will briefly comment on this possibility later on.  $H_{DG}$ coincides with our general $H$ in (\ref{231}) with two different choices of the parameters $\alpha$, $\beta$, $\rho$ and $\mu=\omega\gamma$:

\bea\left\{
\begin{array}{ll}
\mu=s\,e^{i\phi}, \\
\alpha_{\pm}=i\,e^{-i\phi}\left[\frac{r\,\sin(\theta)}{s}\mp\sqrt{\left(\frac{r\,\sin(\theta)}{s}\right)^2-\frac{t}{s}}\right],\\
\beta_{\pm}=i\,e^{-i\phi}\left[\frac{r\,\sin(\theta)}{s}\pm\sqrt{\left(\frac{r\,\sin(\theta)}{s}\right)^2-\frac{t}{s}}\right],\\
\rho_{\pm}=r\,e^{-i\,\theta}+i\,s\left[\frac{r\,\sin(\theta)}{s}\pm\sqrt{\left(\frac{r\,\sin(\theta)}{s}\right)^2-\frac{t}{s}}\right].
\end{array}\label{m2}
\right.
\ena
Moreover, the related values of $\omega_{\pm}$ and $\gamma_{\pm}$ can be deduced by recalling that, in general,
$\gamma=\alpha_{12}\beta_{11}-\alpha_{11}\beta_{12}=\alpha_{12}\beta_{12}(\beta-\alpha)$,  $-\gamma^2=\alpha_{12}\beta_{12}$ and that $(\alpha-\beta)\gamma=1$. Then we deduce that, whenever $\left(\frac{r\,\sin(\theta)}{s}\right)^2\neq\frac{t}{s}$,
\be
\alpha_{12}\beta_{12}=\frac{e^{2i\Phi}}{4\left[\left(\frac{r\,\sin(\theta)}{s}\right)^2-\frac{t}{s}\right]},
\label{m2bis}\en
so that, with a particular choice of the square root,
\be
\gamma_{\pm}=\frac{\pm\,i\,e^{i\phi}}{2\sqrt{\left[\left(\frac{r\,\sin(\theta)}{s}\right)^2-\frac{t}{s}\right]}},
\label{m3}\en
and therefore
\be
\omega_{\pm}=\frac{s\,e^{i\phi}}{\gamma_{\pm}}=\mp\,2\,i\,s \sqrt{\left[\left(\frac{r\,\sin(\theta)}{s}\right)^2-\frac{t}{s}\right]}.
\label{m3b}\en
These results show that, if $\left(\frac{r\,\sin(\theta)}{s}\right)^2\neq\frac{t}{s}$, we can \underline{always} recover a pseudo-fermionic structure for $H_{DG}$, so that all the results deduced and listed previously hold true for this model. The situation changes drastically when $\left(\frac{r\,\sin(\theta)}{s}\right)^2=\frac{t}{s}$. In this case, in fact, $\gamma_{\pm}=0$ necessarily, so that (\ref{230}) cannot be satisfied: in this case no PFs can appear. This is intriguingly related to the existence of EPs in the model, since under this condition the two eigenvalues $E_{\pm}=r\cos(\theta)\pm\sqrt{st-r^2\sin^2(\theta)}$ of $H_{DG}$ coalesce: $E_+=E_-=r\cos(\theta)$.
We also would like to notice that, since $s\in\Bbb R$, $\omega_{\pm}$ are real only if $\left(\frac{r\,\sin(\theta)}{s}\right)^2<\frac{t}{s}$ ({\em unbroken phase}). On the other hand, if $\left(\frac{r\,\sin(\theta)}{s}\right)^2>\frac{t}{s}$, $\omega_+$ and $\omega_-$ are purely imaginary, and one is the adjoint of the other ({\em broken phase}).

\vspace{2mm}

For completeness, we specialize here the relevant quantities deduced previously. In particular,
the eigenvectors of $N$ and $N^{\dagger}$ are given as in (\ref{eigen1}) and (\ref{eigen2}):
\bea
\varphi_0^{(\pm)}=N_\varphi\left(
                     \begin{array}{c}
                       1 \\
                       -\alpha_{\pm} \\
                     \end{array}
                   \right),\qquad \varphi_1^{(\pm)}=b\varphi_0^{(\pm)}=\frac{\gamma_{\pm} N_\varphi}{\alpha_{12}}\left(
                     \begin{array}{c}
                       1 \\
                       -\beta_{\pm} \\
                     \end{array}
                   \right),\label{eigen1bis}\ena
 and                  \bea\Psi_0^{(\pm)}=N_\Psi\left(
                     \begin{array}{c}
                       1 \\
                       \overline{\beta_{\pm}}^{-1} \\
                     \end{array}
                   \right),\quad \Psi_1^{(\pm)}=a^\dagger\Psi_0^{(\pm)}=\frac{\overline{\gamma_{\pm}}\, N_\Psi}{\overline{\beta_{11}}}\left(
                     \begin{array}{c}
                       \overline{\alpha_{\pm}} \\
                       1 \\
                     \end{array}
                   \right).
\label{eigen2bis}\ena
 The lowering and raising operators are also {\em doubled}:
\be
a_{\pm}=\alpha_{12}\left(
       \begin{array}{cc}
         \alpha_{\pm} & 1 \\
         -\alpha_{\pm}^2 & -\alpha_{\pm} \\
       \end{array}
     \right),\qquad b_{\pm}=\beta_{12}\left(
       \begin{array}{cc}
         \beta_{\pm} & 1 \\
         -\beta_{\pm}^2 & -\beta_{\pm} \\
       \end{array}
     \right),
\label{m4}\en
as well as the operators $S_{\varphi}^{(\pm)}$ and $S_{\Psi}^{(\pm)}$, which can be deduced by (\ref{add1}) and (\ref{add2})
specializing the form of the parameters as in (\ref{m2}), (\ref{m2bis}), (\ref{m3}) {and  writing the following values of  $\alpha_{12}$ and $\beta_{11}$
used also to recover the conditions in
\eqref{m2}:
\bea\left\{
\begin{array}{ll}
\alpha_{12}=\frac{2\alpha_{11}\mu}{\mp2is\sqrt{\left[\left(\frac{r\sin(\theta)}{s}\right)^2-\frac{t}{s}\right]}+2ir\sin(\theta)},\\
\beta_{11}=\frac{st}{4\left(st-r^2\sin^2(\theta)\right)\alpha_{11}}.
\end{array}
\right.
\ena
}
Therefore
\be
S_\varphi^{(\pm)}=|N_\varphi|^2\left(
            \begin{array}{cc}
             1+\frac{1}{4}\left|\frac{s \left(x_{rr}^{\pm}\right)}{\alpha_{11} \mu \sqrt{x_r} }\right|^2&   \frac{-i e^{i \phi }}{16} \overline{\left(4\left|\frac{s\left(x_{rr}^{\pm}\right)^2}{\alpha_{11}\mu  \sqrt{x_r}}\right|{}^2x_{rr}^{\mp}+16x_{rr}^{\pm}\right)} \\
               \frac{i e^{-i \phi }}{16} \left(4\left|\frac{s\left(x_{rr}^{\pm}\right)^2}{\alpha_{11}\mu  \sqrt{x_r}}\right|^2x_{rr}^{\mp}+16x_{rr}^{\pm}\right) & \left|x_{rr}^{\pm}\right|^2+\frac{1}{4}\left|\frac{s\left(x_{rr}^{\pm}\right)^2}{\alpha_{11} \mu  \sqrt{x_r}}\right|^2 \\
            \end{array}
          \right)
\label{add1b}\en
and
\be
S_\Psi^{(\pm)}=|N_\Psi|^2\left(
            \begin{array}{cc}
              1+4 \left|\frac{s\alpha_{11} x_{rr}^{\pm}\sqrt{x_r}}{t }\right|^2 & -i e^{+i \phi } \left(\frac{1}{x_{rr}^{\mp}}+4 \overline{x_{rr}^{\pm}} \left|\frac{s\alpha_{11} x_{rr}^{\pm}\sqrt{x_r}}{t }\right|^2\right) \\
            i e^{-i \phi } \left(\frac{1}{\overline{x_{rr}^{\mp}}}+4 x_{rr}^{\pm} \left|\frac{s\alpha_{11} x_{rr}^{\pm}\sqrt{x_r}}{t }\right|^2\right) &\frac{1}{\left|x_{rr}^{\mp}\right|^2}+4 \left|\frac{s\alpha_{11} x_{rr}^{\pm}\sqrt{x_r}}{t }\right|^2 \\
            \end{array}
          \right),
\label{add2b}\en
where we have introduced  $x_r=\left(\frac{r\sin(\theta)}{s}\right)^2-\frac{t}{s}$ and
$x_{rr}^{\pm}=\frac{r\sin(\theta)}{s}\mp\sqrt{x_r}$.

Needless to say, $S_{\varphi}^{(\pm)}$ and $S_{\Psi}^{(\pm)}$ have all the properties we have discussed in Section II.1, and in particular they admit square roots
${S_{\varphi}^{(\pm)}}^{1/2}$ and ${S_{\Psi}^{(\pm)}}^{1/2}$ as in \eqref{add12}-\eqref{add22}. For concreteness sake,
we consider the following particular choice of the parameters of $H_{DG}$: \\
$$
r=1,s=0.5,t=1,\theta=\phi=\pi/6,
$$
and we restrict here to the ”-” choice, fixing also $\alpha_{11}=1$.
Then, our operators look like
\beano
S_\varphi^{(-)}&=&|N_\varphi|^2\left(
            \begin{array}{cc}
             \frac{1}{2} & -0.317+1.549 i\\
             -0.317-1.549i & 3
            \end{array}
          \right),\\
          S_\Psi^{(-)}&=&\frac{1}{2|N_\varphi|^2}\left(
            \begin{array}{cc}
             3 & 0.317-1.549 i\\
             0.317+1.549i & \frac{1}{2}
            \end{array}
          \right)
\enano
and
\beano
{S_{\varphi}^{(-)}}^{1/2}&=&|N_\varphi|\left(
            \begin{array}{cc}
             1.076 & -0.117 + 0.572 i\\
             -0.117- 0.572I & 1.63
            \end{array}
          \right),\\
          {S_{\Psi}^{(-)}}^{1/2}&=&\frac{\sqrt{2}}{2|N_\varphi|}\left(
            \begin{array}{cc}
             1.63 & 0.117+ 0.572i\\
            0.117+ 0.572i & 1.076
            \end{array}
          \right)
\enano

 and we get $$h_{DG}={S_{\Psi}^{(-)}}^{1/2} H_{DG}\,{S_{\varphi}^{(-)}}^{1/2}=\left(
            \begin{array}{cc}
             0.832 & 0.393+ 0.306\,i\\
            0.393- 0.306\,i & 0.9
            \end{array}
          \right),$$
 which is the {\em self-adjoint counterpart} of the hamiltonian $H_{DG}=\left(
            \begin{array}{cc}
             \frac{1}{2}(\sqrt{3}+i) & \frac{1}{4}(\sqrt{3}+i)\\
            \frac{1}{2}(\sqrt{3}-i) & \frac{1}{2}(\sqrt{3}-i)
            \end{array}
          \right)$.

\vspace{2mm}

 {\bf Remark:--} Of course we can obtain the self-adjoint hamiltonian $h_{DG}$ only because $\rho$ and $\omega$ are reals. For the
 particular values of the parameters in $H_{DG}$ considered here we obtain $\rho=1.366$ and $\omega=-1$.

\subsubsection{A particular choice of parameters}

It is interesting to recall that, taking $\phi=0$ and $s=t$ in $H_{DG}$ we recover the hamiltonian
$$
H_{part}=\left(
  \begin{array}{cc}
    r\, e^{i\theta} & s \\
    s & r\, e^{-i\theta} \\
  \end{array}
\right),
$$
considered for instance in \cite{ben1}. Our previous formulas specialize here in an obvious way. In this case, in particular, EPs are recovered for $\frac{r\sin(\theta)}{s}=\pm 1$. Also, $$\omega_{\pm}=\mp2\,i\,s\sqrt{\left(\frac{r\sin(\theta)}{s}\right)^2-1}$$ is real only if $\left(\frac{r\sin(\theta)}{s}\right)^2<1$. Otherwise $\omega_+$ and $\omega_-$ are purely imaginary, and one is the adjoint of the other. EPs appear when $\frac{r\sin(\theta)}{s}=\pm 1$, and in this case PFs are absent.

\subsection{An hamiltonian by Gilary, Mailybaev and Moiseyev}

This hamiltonian was introduced quite recently in \cite{gmm}, and can be rewritten as
\bea
H_{GMM}=\left(
  \begin{array}{cc}
    \epsilon_1-i\Gamma_1 & \nu_0 \\
    \nu_0 &  \epsilon_2-i\Gamma_2 \\
  \end{array}
\right),
\ena
where $\Gamma_1$ and $\Gamma_2$ are positive quantities, $\epsilon_1$ and $\epsilon_2$ are reals, and $\nu_0$ is complex-valued. It is a simple exercise to show that $H_{GMM}$ can be written as in (\ref{231}) with the following identification:
\bea\left\{
\begin{array}{ll}
\omega\gamma=\nu_0, \\
\alpha_{\pm}=\frac{1}{2\nu_0}\left(-\Delta\epsilon+i\Delta\Gamma\mp\sqrt{(-\Delta\epsilon+i\Delta\Gamma)^2+4\nu_0^2}\right),\\
\beta_{\pm}=\frac{1}{2\nu_0}\left(-\Delta\epsilon+i\Delta\Gamma\pm\sqrt{(-\Delta\epsilon+i\Delta\Gamma)^2+4\nu_0^2}\right),\\
\rho_{\pm}=\frac{1}{2}\left(\epsilon-i\Gamma\pm \sqrt{(-\Delta\epsilon+i\Delta\Gamma)^2+4\nu_0^2}\right),
\end{array}
\right.
\ena
where $\Delta\epsilon=\epsilon_2-\epsilon_1$, $\Delta\Gamma=\Gamma_2-\Gamma_1$, $\epsilon=\epsilon_2+\epsilon_1$ and $\Gamma=\Gamma_2+\Gamma_1$. Since $\gamma_{\pm}=\alpha_{12}\beta_{12}(\beta_{\pm}-\alpha_{\pm})$ and $\gamma_{\pm}^2=-\alpha_{12}\beta_{12}$, we find that, whenever $\alpha_{\pm}\neq\beta_{\pm}$, taking
$$
\alpha_{12}\beta_{12}=\frac{-\nu_0^2}{(-\Delta\epsilon+i\Delta\Gamma)^2+4\nu_0^2},
$$
 the pseudo-fermionic main condition is satisfied: $H_{GMM}$ admits a pseudo-fermionic representation. On the other hand, this is not possible if $\alpha_{\pm}=\beta_{\pm}$, which is true when $(-\Delta\epsilon+i\Delta\Gamma)^2=-4\nu_0^2$. Looking at the eigenvalues of $H_{GMM}$, this is exactly the condition which makes its two eigenvalues to coalesce. In this case we have $E_1=E_2=\frac{1}{2}(\epsilon-i\Gamma)$.

The explicit expressions for the relevant eigenvectors and operators can be deduced, as usual, from (\ref{eigen1}), (\ref{eigen2}), (\ref{add1}), (\ref{add2}) and (\ref{m4}).

\subsection{An example by Mostafazadeh and \"Ozcelik}
The model we consider now is different from those above because of the absence of EPs. Then, as we will see, it will always be possible to have PFs for all possible values of the parameters of the model.

The hamiltonian considered in \cite{most2006} is
\bea
H_{MO}=E\left(
  \begin{array}{cc}
    \cos{\theta} & \,e^{-i\phi}\sin(\theta) \\
    \,e^{i\phi}\sin(\theta) &  -\cos{\theta} \\
  \end{array}
\right),
\label{m5}
\ena
where $\theta,\phi \in \mathbb{C} $, $\Re(\theta)\in [0,\pi)$, and $\Re(\phi)\in [0,\pi)$. For obvious reasons we restrict to $E\neq0$ and to $\theta\neq0$. We can deduce two different set of values of $\alpha$, $\beta$, etc. for $H$ in (\ref{231}) such that the two hamiltonians coincide. These choices are
\bea\left\{
\begin{array}{ll}
\mu=E\,\sin(\theta)\,e^{i\phi}, \\
\alpha_{\pm}=\frac{e^{i\phi}}{\sin(\theta)}\left(\cos(\theta)\mp 1\right),\\
\beta_{\pm}=\frac{e^{i\phi}}{\sin(\theta)}\left(\cos(\theta)\pm 1\right),\\
\rho_{\pm}=\pm E.
\end{array}\label{m6}
\right.
\ena
The corresponding pseudo-fermionic operators look  as those in (\ref{m4}), with
$$
\alpha_{12}\beta_{12}=-\,\frac{1}{4}\,\sin^2(\theta)\,e^{-2i\phi}.
$$
Also, there exists no possible condition which makes $\gamma_{\pm}=\alpha_{12}\beta_{12}(\beta_{\pm}-\alpha_{\pm})=0$: contrarily to what happens for $H_{DG}$ and for $H_{GMM}$, this model always allow a pseudo-fermionic description. The eigenvectors of $N$ and $N^\dagger$ are
\beano
\varphi_0^{(\pm)}=N_\varphi\left(
\begin{array}{c}
 1 \\
 \\
 \frac{e^{i\phi}}{\sin(\theta)}\left(\cos(\theta)\mp 1\right)
\end{array}
\right)\quad
\varphi_1^{(\pm)}=\mp N_\varphi\frac{  \cos (\theta )-1}{2\alpha_{11}}\left(
\begin{array}{c}
 -1 \\
 \\
 \frac{e^{i\phi}}{\sin(\theta)}\left(\cos(\theta)\pm 1\right)
\end{array}
\right),\\
\\
\\
\Psi_0^{(\pm)}=N_\Psi\left(
\begin{array}{c}
 1 \\
 \\
 \overline{\left(\frac{e^{i\phi}}{\sin(\theta)}\left(\cos(\theta)\pm 1\right)\right)^{-1}}
\end{array}
\right)\quad
\Psi_1^{(\pm)}=\mp2N_\Psi\frac{ \overline{\alpha_{11} e^{i\phi} }}{\overline{\sin(\theta)}} \left(
\begin{array}{c}
\overline{\frac{e^{i\phi}}{\sin(\theta)}\left(\cos(\theta)\mp 1\right)} \\
 \\
 1
\end{array}
\right)
\enano

In particular, restricting here to the "-" choice, we find that $H_{MO}\varphi_0^{(-)}=-E\varphi_0^{(-)}$, which means that $\epsilon_0^{(-)}=-E$. Moreover, since $\epsilon_1^{(-)}=\epsilon_0^{(-)}+\omega_-$, and since $\omega_-=\frac{\mu}{\gamma_-}=2E$, we deduce that $\epsilon_1^{(-)}=E$. Notice that we have used here
$$
\gamma_-=\alpha_{12}\beta_{12}(\beta_--\alpha_-)=\frac{1}{2}\,\sin(\theta)\,e^{-i\phi}.
$$
The intertwining operators $S_\varphi^{(-)}$ and $S_\Psi^{(-)}$ look now

\beano
S_\varphi^{(-)}&=&|N_\varphi|^2\left(
\begin{array}{cc}
  1+\left|\frac{\cos (\theta )-1}{2\alpha_{11}}\right|^2 & e^{-i \overline{\phi} } \left(\overline{\frac{(1-\cos (\theta)) \left|\frac{\cos (\theta )-1}{\alpha_{11}}\right|^2-\cos(\theta )-1}{4\sin(\theta )}}\right)\\
 e^{i \phi } \left(\frac{(1-\cos (\theta)) \left|\frac{\cos (\theta )-1}{\alpha_{11}}\right|^2-\cos(\theta )-1}{4\sin(\theta )}\right)  & e^{-2 \Im(\phi )}  \left( \left|\frac{(\cos (\theta )-1)^2 }{2\alpha_{11}\sin (\theta )}\right|^2+\left|\frac{\cos (\theta )+1}{\sin (\theta )}\right|^2\right)
\end{array}
          \right)\\
          S_\Psi^{(-)}&=&|N_\Psi|^2\left(
\begin{array}{cc}
\left|\frac{\alpha_{11}}{ \sin ^2\left(\frac{\theta }{2}\right)}\right|^2+1 &  e^{-i \phi }  \left(\overline{\left|\frac{2\alpha_{11}}{ \sin (\theta )}\right|^2  \frac{\cos (\theta )+1}{\sin(\theta)}- \cot \left(\frac{\overline{\theta}}{2}\right)}\right) \\
 e^{i \overline{\phi} }  \left(\left|\frac{2\alpha_{11}}{ \sin (\theta )}\right|^2  \frac{\cos (\theta )+1}{\sin(\theta)}- \cot \left(\frac{\overline{\theta}}{2}\right)\right) & e^{2 \Im(\phi )}  \left(\left|\frac{2\alpha_{11}}{\sin (\theta )} \right|^2+\left|\frac{\sin(\theta)}{\cos (\theta )-1}\right|^2\right)
\end{array}
          \right).
\enano

Moreover if we fix the parameters $\theta=\frac{\pi}{3}+\frac{i}{2},\phi=\frac{\pi}{4}-i,E=1$ in $H_{MO}$ and $\alpha_{11}=1$, we obtain the following representation of
$S_\varphi^{1/2(-)}$ and $S_\Psi^{1/2(-)}$:

\beano
S_\varphi^{1/2(-)}&=&|N_\varphi|\left(
\begin{array}{cc}
 1.076 &-0.709 - 0.005i\\
 -0.709 + 0.005i & 4.532
 \end{array}
          \right)\\
          S_\Psi^{1/2(-)}&=&\frac{1}{|N_\varphi|}\left(
\begin{array}{cc}
1.035 &0.162+0.001i \\
0.162-0.001i &0.245
\end{array}
          \right).
\enano

 The {\em self-adjoint counterpart} of the hamiltonian $H_{M0}$ is
 $$h_{MO}=S_\Psi^{1/2 (-)} H_{MO}\,S_\varphi^{1/2 (-)}=\left(
            \begin{array}{cc}
             0.695 & 0.523- 0.492i\\
            0.523+ 0.492i & -0.695
            \end{array}
          \right).$$

\subsection{A {\em relativistic} example}
We now briefly consider the hamiltonian introduced in \cite{mandal1} and further considered in \cite{ghatak}:
\bea
H_{rel}=\left(
  \begin{array}{cc}
    mc^2 & cp_x+v \\
    cp_x-v &  -mc^2 \\
  \end{array}
\right).
\ena
Here we are assuming that $m$, $v$, $c$ and $p_x$ are all real quantities. If $c^2p_x^2\neq v^2$ $H_{rel}$ can be seen as a particular case of the hamiltonian $H_{MO}$, fixing first
$
\theta=\arctan\left(\frac{c^2p_x^2-v^2}{m^2c^4}\right)$, then taking $E=\frac{mc^2}{\cos(\theta)}$ and finally $\phi=\arccos\left(\frac{cp_x}{E\sin(\theta)}\right)$.
Something interesting happens if $c^2p_x^2= v^2\neq0$. In this case it is easy to check that $H_{rel}$ and $H_{MO}$ are different for any possible choice of the parameters. This is because, while only one non-diagonal matrix element in $H_{rel}$ can be different from zero, the analogous elements in $H_{MO}$ are both zero or both not zero.  Hence the two models, in this case, are really different. However, it is still possible that $H_{rel}$ coincides with $H$ in (\ref{231}). And in fact we find that this is so if $cp_x=v$. In this case, it is enough to fix $\omega\gamma=2v$, $\alpha=0$, $\beta=\frac{mc^2}{v}$ and $\rho=mc^2$ or $\omega\gamma=2v$, $\alpha=\frac{mc^2}{v}$, $\beta=0$ and $\rho=-mc^2$. On the other hand, because of the asymmetry between the (1,2) and the (2,1) elements in $H$, there is no such a possibility if $cp_x=-v$: in this case, PFs are absent.

If $cp_x\neq -v$ the eigenvectors of $H_{rel}$, its expression in terms of pseudo-fermionic operators, the intertwining operators,  can all be deduced adapting our general results to the present situation.

\section{Conclusions}
 We have shown how the general setting of PFs can be used in the analysis of different physical systems introduced along the years in connections with pseudo-hermitian quantum mechanics. The procedure proposed here, other than being rather general and, in our opinion, useful for many other models, provide a set of simple rules and useful results linked to the anti-commutation rules in (\ref{220}). It could be worth mentioning that our analysis does not include all the two-by-two matrices introduced along the years in our context. For instance, in \cite{kjs}, other examples are given, even in higher dimensions. However, the hamiltonian $$H_{JSM}=\left(
                                                                                          \begin{array}{cc}
                                                                                            a & i\,b \\
                                                                                            i\,b & -a \\
                                                                                          \end{array}
                                                                                        \right),
 $$ mentioned in \cite{kjs}, where $a\in{\Bbb R}$ and $b\in{\Bbb R}\setminus\{0\}$, is a particular case of $H_{MO}$: we just have to take $\phi=\frac{\pi}{2}$, and then relate $E$ and $\theta$ to $a$ and $b$.

In our opinion, it is also interesting to stress that the existence of pseudo-fermionic operators appears to be deeply related to the existence of EPs: in fact, in all the models considered here, a lack of validity of (\ref{220}) implies coalescence of eigenvalues. This is expected, since a pseudo-fermionic structure is intrinsically connected with the existence of non coincident eigenvalues. We believe this nice and simple result can be extended to more pseudo-fermionic modes (i.e. to Hilbert spaces with dimension $2^N$, for some natural $N$) and to the much more complicated situation of pseudo-bosons, where (\ref{220}) are replaced by a deformed version of canonical commutation rules, \cite{bag2013}. This will be part of our future analysis.

\section*{Acknowledgements}

The authors acknowledges financial support by the MIUR. F. B. is also grateful to INFN for support.


\begin{thebibliography}{99}

\bibitem{bagpf} F. Bagarello, {\em Linear pseudo-fermions}, J.  Phys. A, {\bf 45}, 444002, (2012)


\bibitem{bagpf2}  F. Bagarello, {\em Damping and Pseudo-fermions},  J. Math. Phys.,  {\bf 54}, 023509, (2013)


\bibitem{bagpantano} F. Bagarello, G. Pantano, {\em Pseudo-fermions in an electronic loss-gain circuit},  IJTP, DOI: 10.1007/s10773-013-1769-y

\bibitem{ben3} C. M. Bender, S. P. Klevansky, {\em PT-Symmetric Representations of Fermionic Algebras}, Phys. Rev. A, {\bf 84}, 024102 (2011)

\bibitem{das} A. Das, L. Greenwood {\em An alternative construction of the positive inner product for pseudo-Hermitian Hamiltonians: examples}, J. Math. Phys., {\bf 51}, Issue 4, 042103 (2010)
\bibitem{deng} J.W Deng, U. Gunther, Q.H Wang{General $\Pc \mathcal{T}$ -Symmetric Matrices}, arXiv:1212.1861 (2012)




\bibitem{ben1} C. M. Bender, M. V. Berry, A. Mandilara, {\em Generalized PT Symmetry and Real Spectra}, J.  Phys. A, {\bf 35}, L467, (2003)

\bibitem{gmm} I. Gilary, A. A. Mailybaev, N. Moiseyev, {\em Time-asymmetric quantum-state-exchange mechanism}, Phys. Rev. A, {\bf 88}, 010102(R) (2013)


\bibitem{most2006} A. Mostafazadeh, S. \"Ozcelik, {\em Explicit realization of pseudo-hermitian and quasi-hermitian quantum mechanics for two-level systems}, Turk. J. Phys., {\bf 30}, 437-443 (2006)



\bibitem{mandal1}  B. P. Mandal, S. Gupta, {\em Pseudo-hermitian interactions in Dirac theory: examples}, Mod. Phys. Lett. A, {\bf 25}, 1723, (2010)

\bibitem{ghatak} A. Ghatak, B. P. Mandal, {\em Comparison of different approaches of finding the positive definite metric in pseudo-hermitian theories}, Commun. Theor. Phys., {\bf 59}, 533-539, (2013)


\bibitem{bag2013} F. Bagarello, {\em More mathematics on pseudo-bosons},  J. Math. Phys., {\bf 54}, 063512 (2013)

\bibitem{kjs} K. Jones-Smith, H. Mathur, {\em A new class of non-Hermitian quantum hamiltonians with PT symmetry}, Phys. Rev. A, {\bf 82}, 042101 (2010)



\end{thebibliography}
\end{document}